**Modal Geometry Governs Proteoform Dynamics**

James N. Cobley

The University of Dundee, Dundee, Scotland, UK.

**Correspondence**: jcobley001@dundee.ac.uk or j_cobley@yahoo.com

**Abstract**

The fundamental laws governing proteoform dynamics have yet to be formulated. As a result, it is unclear how a specific proteoform—a distinct molecular variant of a protein—dynamically shapes its own future by evolving into new modes that exist only in potential until realised. Here, Modal Geometric Field (MGF) Theory couples real and abstract proteoform transitions through four axioms. Axioms 1–3 (invariant) dictate that only first-order transitions occur on the discrete, volume-invariant, non-symplectic modal manifold. Axiom 4 (mutable) projects the occupancy and shape of a real, instantiated molecule into the modal manifold, generating occupancy-induced curvature. By coupling what is real to what is abstract, curvature—which is always conserved—governs proteoform dynamics by dictating the least-action modal transition. Because curvature distribution renders activation energy relative, barriers are mutable, and entropy emerges inevitably from curvature transport. This unification of energy, entropy, and curvature yields hysteresis, path dependence, fractal self-similarity, and trajectories that oscillate between order and chaos. As a scale-invariant and universal framework, MGF Theory reveals how modal geometry governs proteoform dynamics.

**Key words**: Modal Geometry, Proteoform, Curvature, Dynamics, Fractal.

## Introduction

The real physical geometry of a protein molecule—the structure of a specific biochemical mode (state)—defines both its biological identity and function[1]. Discrete molecular shapes arising from mutations, splice variants, or post-translational modifications (PTMs) establish distinct molecular realities: proteoforms[2–6]. Proteoform dynamics—modal transitions—underpin virtually every biological process[3]. These transitions inherently involve both what exists now—the instantiated proteoform mode—and what could exist—abstract modes[7]. This necessarily demands a formal mathematical coupling between real and abstract modes.

Statistical mechanics offers the closest approximation[8]. It interprets thermodynamic quantities like activation energy as emergent from probability distributions over microstates—linking what is to what could be. Yet this framework assumes a flat, undifferentiated state space: all configurations are equipotential prior to Boltzmann weighting. No geometric structure guides transitions; no curvature deforms the landscape to favour one path over another. The possibility space is merely enumerated, not shaped. What is missing is a field theory in which the real proteoform sculpts the geometry of its own potential futures.

This work introduces Modal Geometric Field (MGF) Theory, a scale-invariant, universal physical basis of proteoform dynamics. Here, the real occupancy of an instantiated proteoform molecule mode is functorially projected into the modal manifold as local curvature at the occupied mode. This curvature couples what is real to what is abstract by governing modal transitions. As demonstrated using cysteine proteoforms[9,10], curvature transport between modes renders activation energy relative, providing a mutable geometric primitive for differentiating otherwise iso-potential modal transitions. Entropy then emerges inevitably from curvature transport. This unification of energy, entropy, and curvature yields hysteresis, path dependence, fractal self-similarity, and trajectories that oscillate between order and chaos.

## Results

### Axiom 1. Binomial Theorem Structures the Modal Manifold

In the binary basis, each cysteine residue is either reduced (0) or oxidised (1)[7]. Per combinatorics[11], the modal manifold—$\mathcal{M}$—for a protein with $R$ cysteines is

$$\mathcal{M} = \{0,1\}^R$$

a Boolean lattice (**Figure 1A**) of cardinality

$$|\mathcal{M}| = 2^R$$

Each stratum ($\mathbb{S}_k$) contains modes of exactly $k$ oxidised cysteine residues, of cardinality

$$|\mathbb{S}_k| = \binom{R}{k}$$

Binomial theorem structures the $\mathcal{M}$, with the integer of each strata matching the binomial coefficients enumerated in Pascals triangle.

**Example**: For one of >1,000 human proteins with three cysteine residues[7] like GAPDH:

$$|\mathcal{M}, (GAPDH)| = 2^3 = 8, (|k_0|, |k_1|, |k_2|, |k_3|) = 1,3,3,1.$$

Concretely,

$$k_0 \; = 000, k_1 = 100, 010{,}001, k_2 = 110, 011{,}101, k_3 = 111$$

Axiom 1 applies to any $R$ integer, and naturally extends to a multinomial lattice, accommodating multiple PTMs from sulfenic acids to disulfide bonds[12]. The multinomial theorem extends axiom 1 to the full proteoform universe (**Supplementary Information**).

**Axiom 2. The Modal Manifold is Volume Invariant.**
Since the $\mathcal{M}$ enumerates every mode, each molecule must occupy exactly one mode at a time. Modes cannot be created or destroyed, so the total volume of the manifold—the probability mass distributed across its modes—remains invariant.

Modal transitions can redistribute occupancy between modes but never alter the total volume. For example, a system comprising molecules with three cysteines always has eight modes (Figure 1B). They persist as a latent geometric structure even when unoccupied. The manifold is non-symplectic, meaning that conservation here does not arise from Hamiltonian mechanics but from the combinatorial structure of the $\mathcal{M}$ itself. Hence, volume invariance is a fundamental property of proteoform dynamics (**Supplementary Information**).

Formally, if $(\rho_x)$ denotes the occupancy of mode $x$. Global conservation requires

$$\frac{d}{dt}\sum_{x\in\mathcal{M}} \rho(x) \; = \; 0$$

and for discrete modal transitions between modes $x$ and $y$, with rates $k_{xy}$ occupancy evolves as

$$\dot{\rho}(x) \; = \sum_{y e N(x)} \left[\rho(y)k_{yx} - \rho(x)k_{xy}\right], \sum_{x} \dot{\rho}(x) = 0$$

**Axiom 3. First-Order Modal Transitions**
Axiom 1 defines every mode the molecule could occupy while axiom 2 ensures they are always conserved. As a result, a molecule—a classical object—must assume one—and only one—of these identities at a time. Hence, the occupancy $\rho(x)$ of an instantiated molecule must equal 1 at the occupied mode and zero elsewhere:

$$\rho(x) \; = \begin{cases} 1, & x = x_i \\ 0, & otherwise \end{cases}$$

As a result, a sequence, such as 000 $\rightarrow 010 \rightarrow 011$, must proceed in a first-order stepwise fashion. These first-order dynamics arise because it is physically impossible for a classical object to occupy two modes at a time—fractional transport of classical unit mass is forbidden. Hence, discrete single step modal transitions between nonadjacent modes (e.g., $000 \rightarrow 011$) are barred.

Formally, nonzero transition probabilities $P(x \rightarrow y)$ must satisfy

$$Hamming(x, y) \; = \; 1$$

The allowed outflow from any mode forms a probabilistic simplex:

$$\sum_{y:Ham(x,y)=1} P(x \to y) = 1, P(x \to z) = 0 \; if \; Ham(x \to z) > 1$$

**Example**: The only allowed neighbours for the 000 mode are {100, 010, 001}. All the other modes lie at hamming distance $> 1$ and have $P = 0$ (**Figure 1C**).

Axiom 3 formalises the primitive constraint that a classical object cannot bifurcate across modes. Multi-site changes decompose uniquely into sequences of Hamming-1 steps, reflecting the separability of modal transitions in space and time (**Supplemental Information**).

Hence, Axioms 1–3 place invariant geometric constraints on proteoform dynamics: all action must proceed in discrete steps on a bounded manifold, restricted to Hamming-1 dynamics.

**Axiom 4. Occupancy Curves the Modal Manifold**

Every molecule belongs to the modal manifold $\mathcal{M}$ (Axiom 1). When a molecule instantiates a mode, it does so with both an identity (molecular occupancy) and a shape (atomic structure). These real attributes are projected functorially into $\mathcal{M}$, deforming its structure and endowing it with intrinsic curvature (**Figure 1D**).

Functorial projection means that every sequence of real molecular updates has a corresponding sequence of geometric updates. Composition is preserved: transitions in the real category induce a lawful algebra of transformations in the abstract category, where occupancy redistribution becomes curvature redistribution and metric deformation. This provides a natural algebra of proteoform dynamics: the manifold itself is reshaped according to conserved rules, ensuring that identity and geometry remain coupled under all transformations.

Formally, if $\mathcal{F}: (\rho, \varphi) \mapsto R$ denotes this projection, then curvature is

$$R(x) = (L_{\varphi}\sigma)(x), \sum_{x \in \mathcal{M}} R(x) = 0$$

where $\sigma \; G(\rho, \varphi)$ encodes occupancy and shape, and $L_{\varphi}$ is a symmetric zero-sum Laplacian.

Hence, curvature cannot be created or destroyed: it is the conserved geometric expression of molecular identity, transported across $\mathcal{M}$ during transitions.

In MGF Theory, curvature therefore differentiates transitions on the modal manifold: it deforms the space, introduces a geometric penalty to the action of energy, and resolves what would otherwise be iso-potential moves. In this way, curvature is the conserved regulator that couples molecular identity to proteoform dynamics, explaining how modal geometry governs proteoform dynamics.

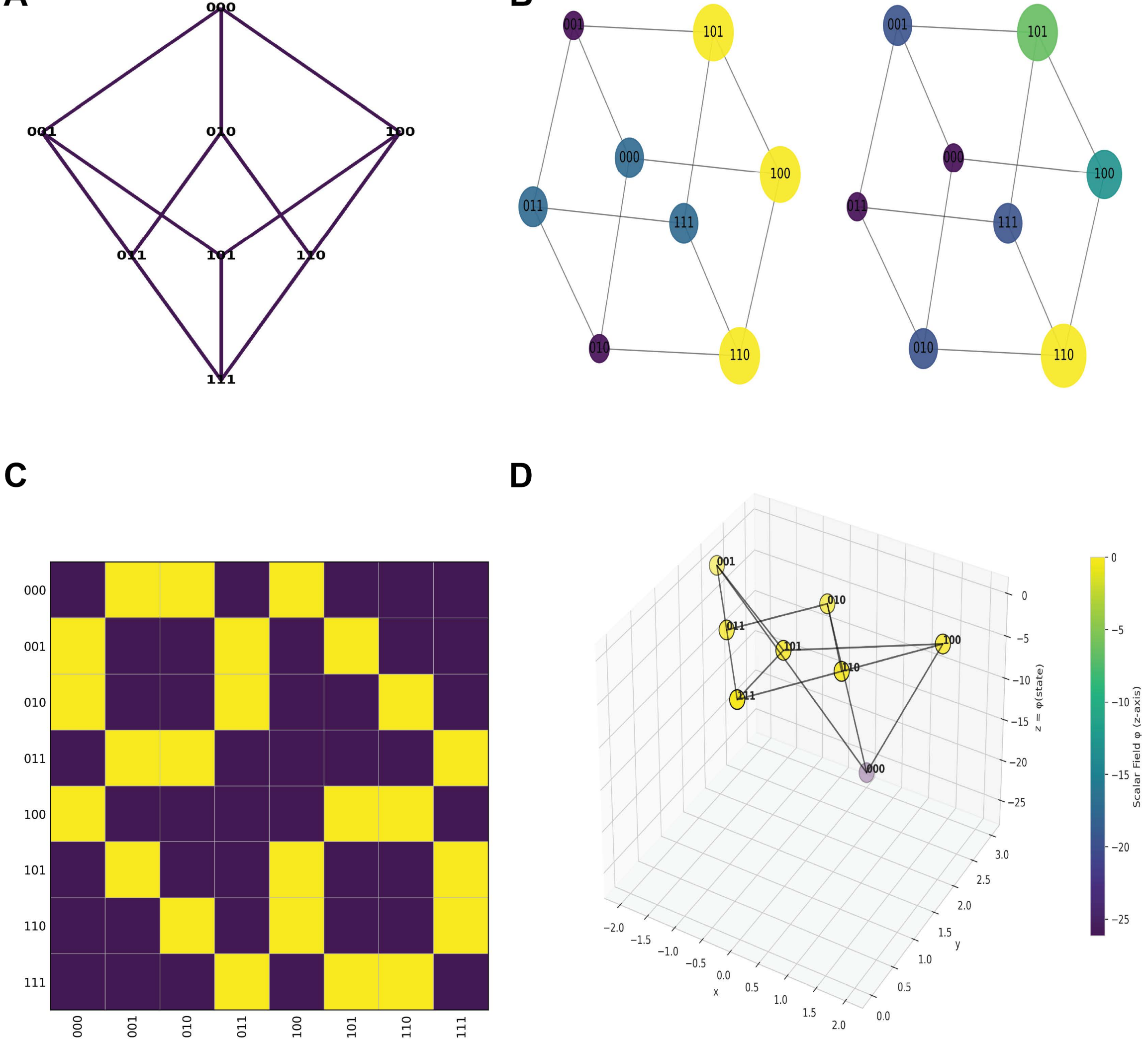

**Figure 1. MGF Theory Axioms. A. Binomial theorem structured (1:3:3:1) Boolean lattice of proteoform modes in the combinatorics-enumerated $R = 3$ series. B. The modal manifold is volume-invariant. No matter how molecules redistribute—depicted as a change in size of the nodes in left vs. right—the volume of the hypercube remains invariant. C. Adjacency matrix of allowed (yellow) and barred (purple) first-order modal transitions. D. Occupancy—at "000"—induced deformation of the modal manifold, generates scalar Ricci curvature.**

## The MGF Theory Field Equation

The MGF Theory unites energy, curvature, and entropy via the modal transition function

$$T_{i\to j} = \underbrace{\Delta E_{i\to j} \cdot exp\,[\rho(x_i) \cdot \Delta H_1]}_{Energy} - \underbrace{R(x_i)}_{Curvature} + \underbrace{\Delta S}_{Entropy}$$

- $T_{i\to j}$: Modal transition function governing the move from mode $i$ to mode $j$.
- $\Delta E_{i\to j}$: Energy cost of the modal transition.
- $\rho(x_i)$: Local occupancy at mode $x_i$ in the discrete, non-symplectic volume invariant manifold ( $\mathcal{M}$, axiom 1-2).
- $\Delta H_1$: First order Hamming-1 transition (axiom 3).
- $R(x_i)$: Discrete Ricci curvature at mode $x_i$ (axiom 4), defining the geometric penalty associated transporting curvature.
- $\Delta S$: Entropy gain from heat, conformational changes, and disorder.

To enforce the conservation of probability across $\mathcal{M}$, one can impose the stationary condition:

$$\sum_{k=0}^{R} P(i \to j) \cdot T_{i \to j} = 0$$

ensuring that transitions respect global volume invariance axiom 2 while admitting the local redistribution of curvature per axiom 4.

Axioms 1–3 define invariant structural geometric constrains: the Boolean hypercube enumerates modes, volume is conserved, and transitions are restricted to first-order moves. Axiom 4 introduces curvature as a mutable regulator: concentrated occupancy creates steep wells that suppress transitions, while distributed occupancy flattens the manifold and facilitates motion. This is analogous to how transporting distributed mass (multi-modal distribution) is easier than concentrated mass (unimodal distribution). This curvature is the discrete analogue of energy basins, projected lawfully from real proteoforms into $\mathcal{M}$. The field equation encodes a modal geometry of proteoform dynamics.

**Modelling MGF Theory**

To model the MGF equation, the stepwise action integral ($\mathcal{A}$) was defined as:

$$\mathcal{A} = \sum_{s=1}^{S} [\Delta S(s) + \lambda_R R(x_s)^2 + \lambda_A \| A(x_s) \|^2]$$

where $\Delta S$ is the entropy gain, $R(x_s)$ the local curvature, and $A(x_s)$ the anisotropy at mode $x_s$.

Per the methods, $10^3$ synthetic priors—each consisting of 100 molecules randomly distributed across the $\mathcal{M} = \{0,1\}^3$ modal Boolean manifold—were evolved for 100 steps under the probabilistic action limited evolution engine (ALIVE) algorithm. The action integral was instantiated along each trajectory, with entropy, curvature, anisotropy, and degeneracy contributions computed at every step and accumulated across runs.

Regression analysis revealed that geometry shaped proteoform dynamics (**Figure 2A**). High curvature suppressed $\mathcal{A}$ ($\beta = -1059.7$, $p < 1e{-94}$), whereas directional anisotropy strongly amplified it ($\beta = +2112.8$, $p < 1e{-88}$). Hence, modal geometry governs proteoform dynamics, constraining or promoting transitions depending on curvature and anisotropy.

A discrete Lie-like operator algebra over $\mathcal{M}$ was developed. Each bitflip defined an operator $M_i$ acting on a given mode $x \in \mathcal{M}$. The resultant algebra is non-abelian: some operators cannot commute. For example, 000 and 111 cannot directly commute per axiom 3.

Finite modal sequences can commute (**Figure 2B**). For example, the sequence $000 \to 100 \to 101 \to 111$ defines a communicative algebraic substructure. Communicative substructures define geodesics—the shortest pathlength between 000 and 111 (**Figure 2C**)—per:

$$|Geodesics_k| = k!$$

yielding six geodesics:

1. $000 \rightarrow 100 \rightarrow 101 \rightarrow 111$
2. $000 \rightarrow 100 \rightarrow 110 \rightarrow 111$
3. $000 \rightarrow 010 \rightarrow 011 \rightarrow 111$
4. $000 \rightarrow 010 \rightarrow 110 \rightarrow 111$
5. $000 \rightarrow 001 \rightarrow 101 \rightarrow 111$
6. $000 \rightarrow 001 \rightarrow 011 \rightarrow 111$

In the simulations, curvature introduced path-dependence across geodesics. Distinct geodesics connecting the same start (000) and end (111) mode produced asymmetric operator orderings (**Figure 2D**). Curvature enforced a non-abelian structure: operations that commute within a geodesic became non-commutative across geodesics. Proteoform dynamics can exhibit path-dependent hysteresis[13].

The nested, recursive self-similar structure of the geodesics (**Figure 2E**) maps to the Sierpiński triangle when Pascal's triangle is reduced modulo 2[14]. While these fractal-like structures evoke order and chaos[15,16], they take on a different meaning in a discrete, volume-invariant manifold. Axiom 1 bounds every Lyapunov exponent[17,18] . In MGF Theory, Ricci Flow[19]

$$\Phi(\rho) = \max_{x, y, \in \mathcal{M}} |R(x;\rho) - R(y;\rho)|$$

behaves as a simple harmonic oscillator generator of $order \rightarrow chaos \rightarrow order$. At the left extreme, the system is maximally ordered. Local curvature is concentrated by all of the molecules occupy one mode (agnostic of mode identity). At the bottom, the system is maximally chaotic. Local curvature is dispersed evenly across all of the modes. No curvature-induced *order* weights modal transitions. Like a thermal system at equilibrium, they are all iso-potential. At the right extreme, the system is reordered—local curvature is reconcentrated by unimodal occupancy.

Modal dynamics span the full oscillator spectrum, admitting different frequencies of cycling between order and chaos. In this view, entropy is the inevitable, direction-agnostic product of curvature spread, quantified by

$$\Phi(\rho) = \sum_{x \in \mathcal{M}} \left(\rho_x - \frac{1}{|\mathcal{M}|}\right)^2$$

Where $\rho_x$ is the normalised occupancy distribution. The bounded evolution of Ricci Flow admits a natural wave mechanic interpretation

- **Amplitude**: defined by modal degeneracy $D(t)$, bounded between one (all molecules in a single mode, maximum order) and $R$ (uniform dispersion, maximum chaos).
- **Frequency**: determined by the periodicity of Ricci spread oscillations. Distinct trajectories generate different frequencies, reflecting asymmetry and geodesic trapping.
- **Phase**: set by operator ordering along geodesics. Non-commuting operator sequences introduce phase shifts, encoding path-dependence.

Across the simulations, the Ricci Flow amplitude of the ensemble trajectories showed that $D(t)$ fluctuates within this strict bound, oscillating between lower-degeneracy ordered states and higher-degeneracy chaotic states (**Figure 2F**). The black mean curve demonstrates the emergent wave-like behaviour of degeneracy, with trajectories distributed around it within one standard deviation.

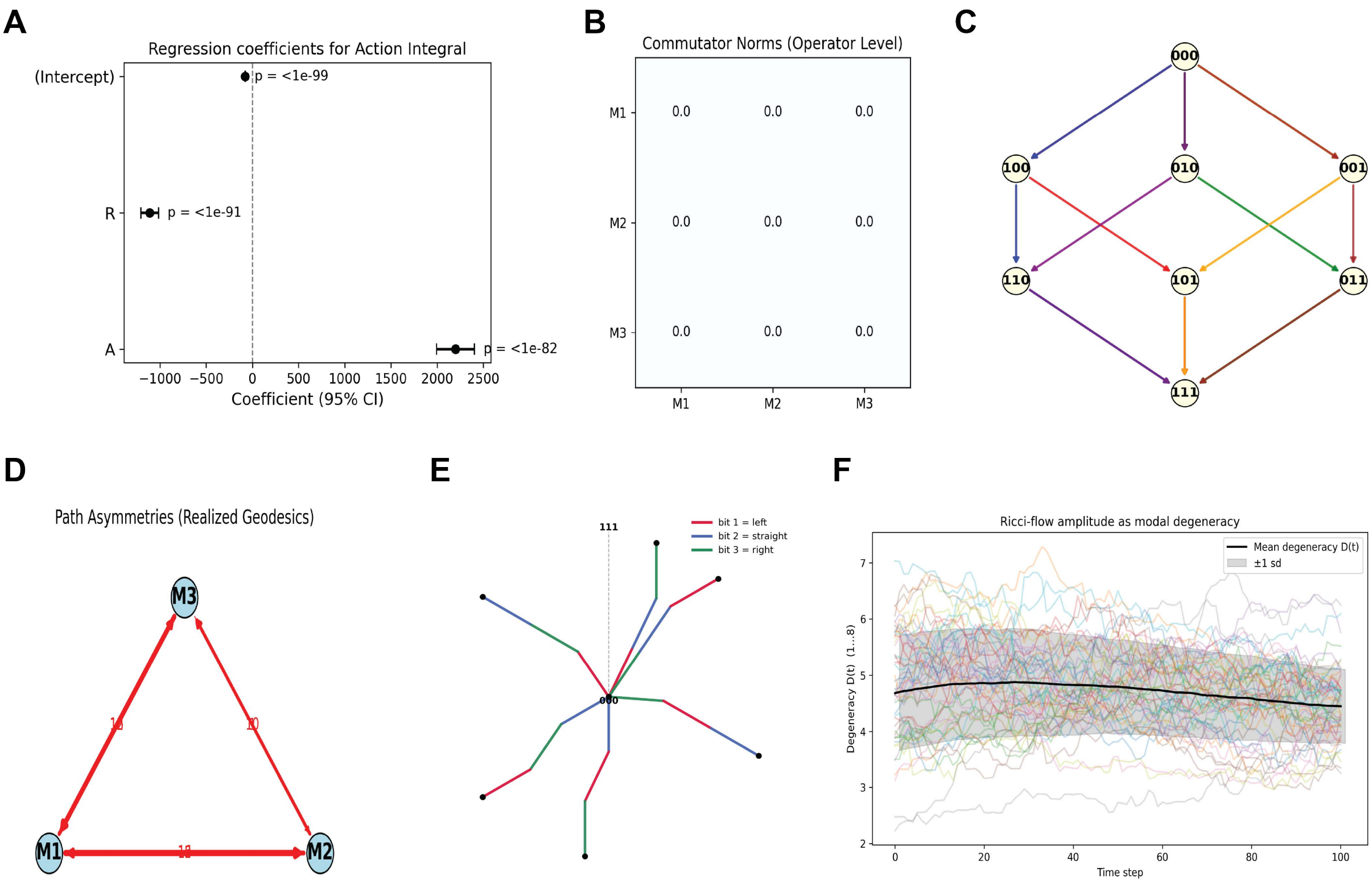

**Figure 2. Modal Geometry Governs Proteoform Dynamics. A. Forest plot of regression coefficients (β, 95% CI) from general linear modelling of the action integral against mean curvature (*R*) and anisotropy (*A*). B. Heatmap of commutator norms between the empirical generators M1,M2,M3, where the numbers map to each cysteine bit. All values are near zero, consistent with symmetric operator action between allowed modes. C. Geodesics on the Boolean manifold. D. Directed graph of observed path motifs. Node size denotes operator identity, arrow thickness encodes motif frequency, and red arrows highlight asymmetries where $M_i \rightarrow M_j \neq M_j \rightarrow M_i$. Asymmetry represents non-commutativity geodesic orderings, even when the operator algebra appears symmetric as in panel C. E. Symbolic representation of the self-similar geodesics. From 000 radiating to 111, oxidation of bit 1, 2 and 3 deviates the trajectory of the line to the left, centre, and right, respectively. F. Ricci-flow amplitude across the simulated trajectories, the black line shows the mean degeneracy, one standard deviation is shown in grey.**

Selected trajectories were selected projected into the complex domain as an example of discrete Mandelbrot-like[20] shapes (**Figure 3**). Symmetric trajectories remained on the geodesic (trap fraction ≈ 0.97, non-commutativity = 0), producing ordered Ricci waves with narrow spectra and nearly holomorphic complex orbits. Trapped trajectories deviate (trap fraction ≈ 0.73, noncommutativity = 0.5), yielding chaotic Ricci waves, broadband spectra, broken holomorphy, and asymmetric operator adjacency.

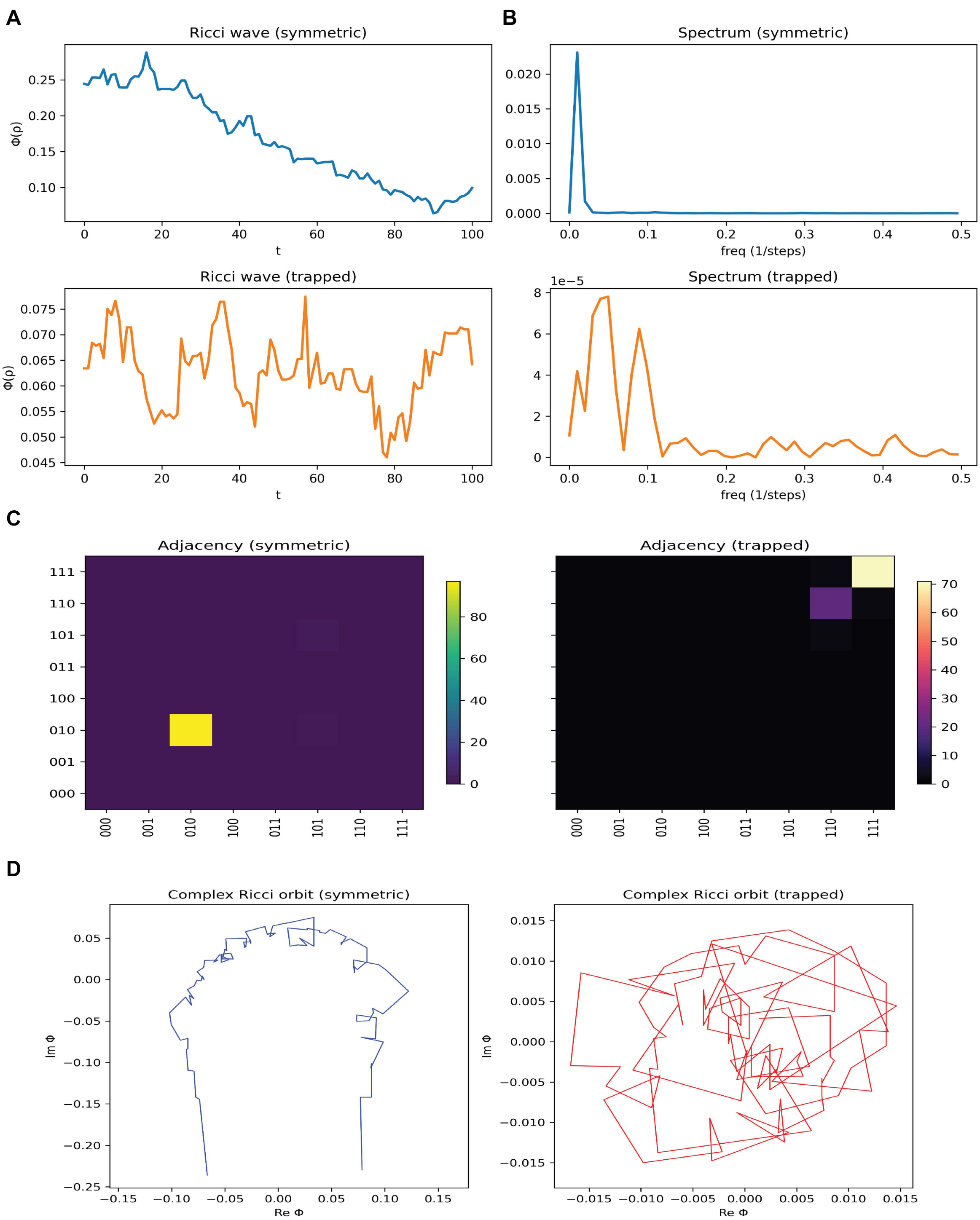

**Figure 3. Ricci-wave dynamics reveal ordered vs trapped trajectories. A-B. Time series of Ricci spread Φ(ρ) and corresponding Fourier spectra for a representative symmetric trajectory (top) and a trapped trajectory (bottom). Symmetric dynamics exhibit smooth oscillations with narrow spectral content, whereas trapped dynamics show irregular fluctuations and broadband spectral power. C. Transition adjacency matrices of the same trajectories. Symmetric flow produces a sparse, geodesic-aligned structure, while trapped flow shows asymmetric clustering of transitions, indicative of non-commutativity. D. Complex projection of Ricci-wave trajectories using the analytic signal. Symmetric trajectories form near-holomorphic orbits in the complex plane, while trapped trajectories generate irregular, non-holomorphic paths with broken symmetry.**

Modelling the MGF Theory field equation demonstrated that the Boolean hypercube constrained trajectories, volume was conserved, and transitions followed first-order moves (axioms 1–3). Curvature acted as a mutable regulator (axiom 4), with concentrated occupancy suppressing transitions and distributed occupancy facilitating motion. The resulting Ricci waves, spectra, and complex orbits demonstrate how curvature, energy, and entropy govern proteoform.

### Conservation of Curvature

While Ricci Flow[19] can smooth the collective curvature of multi-molecule occupancy distribution molecules in $\mathcal{M}$, curvature is conserved at the single-molecule level. Formally, for a single-molecule occupying mode $x$, curvature is defined as the Laplacian of occupancy

$$R(x) = \Delta\rho(x)$$

Direct enumeration across all eight modes in the $R = 3$ series, revealed that while curvature is nonzero at the occupied mode the total curvature is zero

$$\sum_{x \in \mathcal{M}} R(x) = 0$$

The globally "flat" $\mathcal{M}$ is locally deformed by molecular occupancy. In analogy to Noether's theorem[21], the volume-invariance of the discrete non-symplectic $\mathcal{M}$ (axiom 2) guarantees a conservation of curvature law (axiom 4). Curvature cannot be created or destroyed—only transported between neighbouring modes.

Computational enumeration of all single-molecule modal placements and random occupancy distributions (**Figure 4**) yielded $\sum_x R(x) = 0$ within machine tolerance ($< 10^{-12}$). Hence, curvature is conserved.

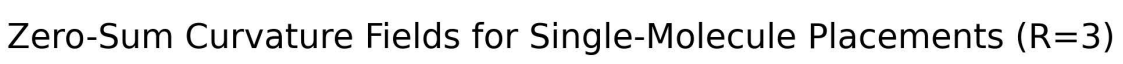

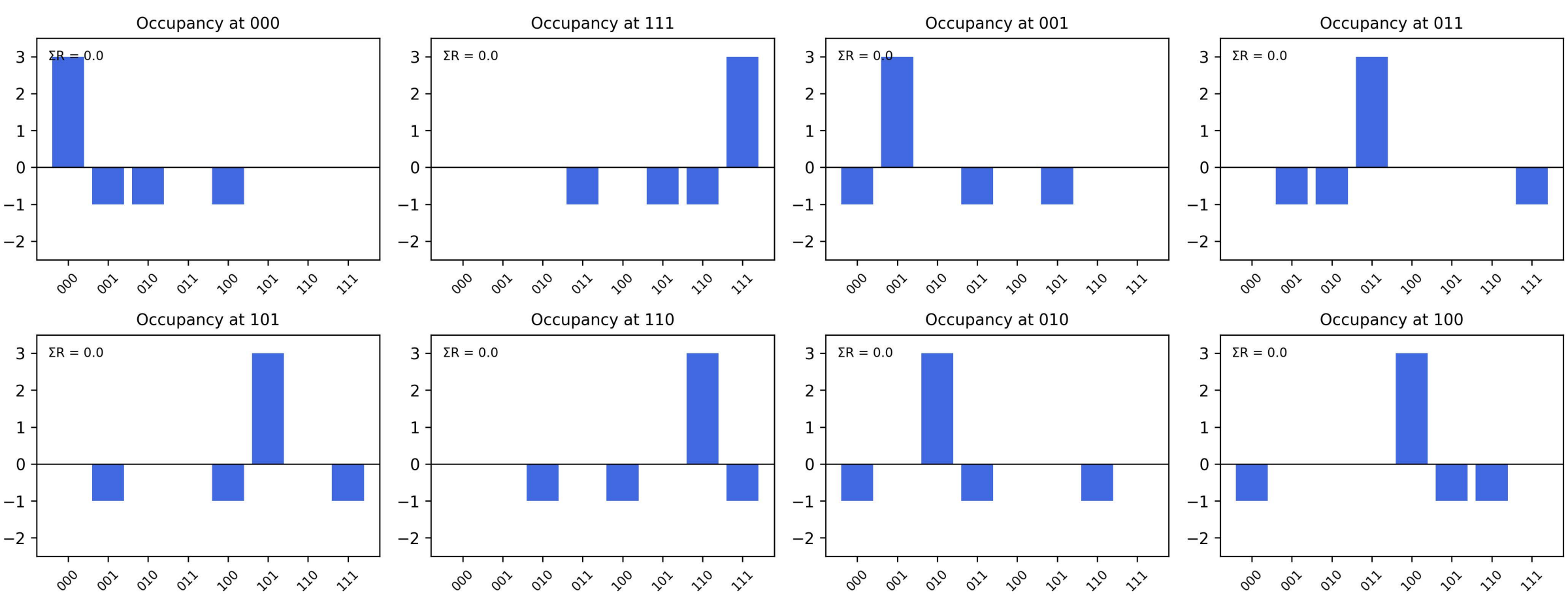

**Figure 4. Conservation of curvature. The bar plot series showing the distribution of curvature across and the total curvature across each instantiated modal occupancy. In this case, the occupied node carries positive curvature (+3), its three Hamming-1 neighbours are compensated with (-1) curvature. Hence, despite the local deformation the global curvature is zero (3 – 3 = 0). The volume-invariant manifold, means that while curvature is mutable—it can be transported—the total curvature must be 0, establishing a discrete Noether-like analogue on the non-symplectic modal manifold.**

**The Geometric Primitive**

A single-molecule instantiated in mode "000" must evolve to a valid neighbour ("100", "010", or "001'), if it is acted on or remain inert—establishing a discrete Einstein-Hilbert least-action analogue of axiom 3:

$$\mathcal{A} = \int R\sqrt{g}\,dv$$

If the "000" mode were acted upon, then "100", "010", and "001" are all iso-potential ($R\ (x) = -1$, **Figure 4**), and, therefore, equally probable ($P = 0.33$). The total curvature of the occupied mode (+3) is evenly partitioned across Hamming-1 neighbours because each bit is currently treated identically. Hence, nothing biases one allowed modal transition from another.

To differentiate otherwise iso-potential modal transitions, a Poisson partial differential equation (PDE) was solved on the atomic structure of human GAPDH ($R$ = 3). For each cysteine bit, the PDE yielded a Dirichlet energy ($E_b$), and a local source energy ($S_b$), whose ratio defined a normalised bitwise projection weight ($w_e$). These values were functorially lifted into the modal Laplacian, yielding

- Cys152(bit1). $E_b = 2.42 \times 10^{-1}, S_b = 1.58 \times 10^{-1}, w_E = 0.388$
- Cys156 (bit2). $E_b = 2.11 \times 10^{-1}, S_b = 1.25 \times 10^{-1}, w_E = 0.351$
- Cys247(bit3). $E_b = 3.02 \times 10^{-1}, S_b = 1.32 \times 10^{-1}, w_E = 0.260$

By contrast persistent topological analysis of the local bitwise graph yielded nearly uniform Morse weights ($w_M \approx 0.333$), consistent with a symmetry-preserving background.

On the modal manifold, Dirichlet energies computed with the instantiated mode—source—fixed at "000" were:

- Energy-channel: $E = 4.62 \times 10^{-1}$
- Morse-channel: $E = 4.53 \times 10^{-1}$
- Fused (average): $E = 4.55 \times 10^{-1}$
- Fused (tropical): $E = 4.54 \times 10^{-1}$

Instantaneous outflow probabilities from a constrained Markov transition model under each channel further confirmed the PDE bias:

- Energy-channel: bit1$> bit2 > bit3$
- Morse-channel: equal.
- Fused (average): bit1$> bit2 > bit3$
- Fused (tropical): bit1$> bit2 > bit3$

The tropical fusion implements the algebra of minimum action, selecting per bit the least-cost contribution between PDE curvature-energy and Morse topology. In this sense, it is a discrete instantiation of the Einstein-Hilbert variational principle on the modal manifold.

Time-evolution on the modal Laplacian confirmed that under the PDE-derived channel the "100" modal transition from "000" is consistently the geometric primitive. Biasing edge weights by the curvature of the PDE, revealed that bit3 had the greatest intrinsic curvature—consistent with resistance to action.

Functorial projection of PDE-derived curvature-energy into the modal manifold broke the degeneracy of iso-potential modal transitions, identifying $000 \rightarrow 100$ as the geometric primitive. This matches empirical evidence[22–24]: Cys152 is experimentally verified as the most oxidation-prone cysteine[25]. The model predicts that the "111" human GAPDH mode[26] arises via preferential action along the $000 \rightarrow 100 \rightarrow 110 \rightarrow 111$ geodesic. Beyond GAPDH, bitwise differentiation of the modal manifold provides a general method of identifying geometric primitives.

**Curvature, Energy, and Entropy**

Since the intrinsic curvature of the $\mathcal{M}$ is invariant (axiom 4), a mode will remain inert—persisting indefinitely ($x_i \rightarrow \infty$) in the absence of forcing—per Newton's first law. External energy ($E$), such as a redox reaction event[27,28], can pay the cost of transporting curvature from $x_i \rightarrow x_j$. Whether expressed as heat dissipation or configurational rearrangement, entropy ($S$) is an inseparable product of energy-coupled curvature transport per Ricci Flow, such that

$$E = x_i \rightarrow x_j + \Delta S, \qquad S(x_i \rightarrow x_j) > \mathbf{0}$$

MGF Theory projects the real occupancy into the discrete $\mathcal{M}$ to calculate curvature

$$\Delta G = \kappa_{modal} |R(x)|,$$

where $R(x) = (L\rho)(x)$ is the scalar curvature induced by occupancy, and $\kappa_{modal}$ is the geometric coupling constant. This relation states that curvature and energy are proportional, in exact analogy to general relativity[29], where curvature is tied to mass-energy via $G_{\mu v} = 8\pi T_{\mu v}$.

In this view, curvature becomes an energetic field property: any transport of modal curvature incurs a quantised energy cost—the activation energy[30] ($E_a$) needed to transport modal curvature. Curvature transport only occurs when the necessary $E_a$ is supplied. As energy is conserved, any surplus is partitioned into entropy and free energy ($\Delta G$)

$$E_{IN} = E_a(x_i \rightarrow x_j) + \mathit{\Delta} S + \mathit{\Delta} G$$

**Numerical example**: For a GAPDH molecule instantiated in "000" the modal manifold, occupancy induced curvature produced a symmetric barrier height of $\approx 33\ kJ \cdot mol^{-1}$ per cysteine bit. By construction, this value is the same for all three bits, since the curvature spectrum is invariant under axiom 4.

Per the geometric primitive, the bitwise energy weights, calculated via the PDE solution of the real atomic structure, were used to differentiate the modal manifold

$$E_a(b) = \kappa_{modal} |K| (1 - \text{In}\, w_E(b))$$

where $|K|$ is the symmetric barrier height ($\approx 33\ kJ \cdot mol^{-1}$). Applied to Cys152, this equation yielded an activation energy of $76\ kJ \cdot mol^{-1}$, which is congruent with $67\ kJ \cdot mol^{-1}$ as computed from empirical kinetic data[31] (see methods).

In MGF Theory, the mutability of curvature—it varies as occupancy changes—makes the activation energy relative. This idea lies at the centre of the field equation where the penalty cost, such as the activation energy, of a modal transition decreases as Ricci Flow dissipates curvature.

**Scale-invariance**
MGF Theory is scale-invariant: it applies to any nonzero $R$ integer. Adapting scripts to accommodate different $R$ integers, revealed that the same axioms and equations applied across the scales. Regardless of how many modes compromise the manifold—its modal degrees of freedom, the laws acting on them remain the same.

**Universality**
Cysteine is no special case. MGF Theory generalises to any PTM on any amino acid (AA). Formally, the mathematics of an $R = 3$ $\mathcal{M}$ are invariant, regardless of whether the three bits represented cysteine redox or tyrosine phosphorylation modes. For example, per axiom 1, there would be eight tyrosine phosphoforms modes: 000, 100, 010, 001, 110, 011, 101, and 111. Like the real to abstract categorical projection, AA-specified PTM morphisms of $R$ can be described by a functor

$$\mathcal{F}: AA + PTM \longrightarrow \mathcal{M}$$

The functor projects any AA-specified PTM-basis, from lysine acetylation to tryptophan oxidation, into the modal manifold. Under any functorial transformation, the axioms of MGF Theory remain invariant. Hence, MGF Theory is universal: It applies to all proteoforms, including every instantiated splice isoform and/or genetic variant.

**Conclusion**
MGF Theory explains how real and abstract modes are geometrically coupled through four axioms, uniting energy, entropy and curvature into a single framework. This formulation is consistent with thermodynamics and statistical mechanics, but crucially extends them geometrically. At this resolution, the globally "flat" modal manifold is locally deformed by occupancy, allowing iso-potential modes to be distinguished via geometric primitives. These primitives define least-action pathways that bias proteoform dynamics toward preferred geodesics, while still permitting alternative but geometrically lawful trajectories.

## Methods

***Source code***
The Julia[32]-scripted source code used to computationally instantiate the mathematics is available online at https://github.com/JamesCobley/Oxi_Shapes/tree/main.

***Modelling the MGF Theory Field Equation***
To model the MGF Theory field equation, the $\mathcal{M}$ was initialised as a 2-dimensional diamond shaped—1:3:3:1—Boolean lattice graph, were the edges defined allowed—first-order—transitions between modes per axiom 1 and 3. To enforce axiom 2, the volume of the modal manifold as the probability of molecular distribution of the molecules was invariant:

$$\sum_{x} P(x) = 1$$

To enforce axiom 4, the $R(x)$ curvature was quantified as the local density imbalance in the $\mathcal{M}$, given by the graph Laplacian:

$$R(x) = \sum_{y \in N(x)} [\rho(y) - \rho(x)]$$

Where the $N(x)$ is the allowed neighbouring coordinates in the $\mathcal{M}$. To encode local geometric heterogeneity, each mode was treated as an anisotropic fibre bundle per:

$$A(x_i) = \frac{1}{|N\ (x_i)|} \sum_{x_j \in N(x_i)} \frac{|R(x_i) - R(x_j)|}{\left\| x_i^{(3D)} - x_j^{(3D)} \right\|}$$

- $x_i \in$ is a mode in the binomial lattice.
- $N(x_i)$ is neighbourhood of $x_i$, defined by a Hamming distance =1.
- $R(x)$ is the local scalar curvature at node $x$, computed as the discrete Laplacian of $\rho(x)$.
- $x_i^{(3D)} = (x_i^x, x_i^y, -\rho(x_i))$ are the 3D coordinates of $x_i$.
- $x_i^{(3D)} - x_j^{(3D)}$ is the Euclidean distance in the curved geometry.
- $A(x_i)$ is the anisotropy at node $x_i$, encoding local directional asymmetry in curvature.

These fibres are wrapped in a self-consistent sheath, representing the global volume invariant manifold. Modal transitions were then simulated per the field equation

$$T_{i \to j} = \underbrace{\Delta E_{i \to j} \cdot exp\,[\rho(x_i) \cdot \Delta H_1]}_{Energy} - \underbrace{R(x_i)}_{Curvature} + \underbrace{\Delta S}_{Entropy}$$

The $10^3$ modal occupancy distributions ($\rho$) were initialised from the random sampling of 100 molecules per run. The ALIVE algorithm stochastically updated occupancies across the $\mathcal{M}$ using curvature-constrained Monte Carlo steps[33,34], with anisotropy acting as a directional penalty/weight. Given the biophysical constraints on energy-coupled redox reactions, the maximum number of moves permitted per step was set to ten. Each 100-step trajectory produced a time-series of $\rho(x)$, $R(x)$, and $A(x)$.

***Action integral***

To compute $\mathcal{A}$, entropy was quantified from symmetry degeneracy ($S_{DEG}$). Each mode at $x$ with hamming weight $k$ was weighted by the binomial degeneracy:

$$S_{DEG}\ (\rho) = \sum_i \rho(x_i) log \frac{1}{\binom{R}{k}}$$

Where $R$ is the residue integer and $\binom{R}{k}$ is the degeneracy of that oxidation integer (e.g., $k = 1$, degeneracy = 3). The cumulative $\mathcal{A}$ along a trajectory was defined as the weighted sum of three contributions per time step:

$$\mathcal{A} = \sum_t \left[ \alpha_{mass} \cdot \sum |\Delta_\rho| + \alpha_{geom} \cdot (\langle R^2 \rangle + \langle \|A^2\| \rangle) + \alpha_{entropy} \cdot S_{deg}(\rho) \right]$$

Where $\Delta\rho$ is the flux, $R$ the local curvature, and $A$ the local anisotropy. The parameters $(\alpha_{mass,}\alpha_{geom,}\alpha_{entropy})$were set to 0.01, 0.1, and 0.1, respectively. After computing $\mathcal{A}$ along all of the trajectories, statistical regression analysis was performed to determine the influence of curvature and anisotropy on action.

***Discrete Algebra***

A discrete operator algebra was constructed directly from the simulated proteoform trajectories. For each trajectory, stepwise transitions were parsed and encoded as triples (to,from,$b$), where $b$ indexes the bit flipped. Each bit-flip thereby defined an operator $M_b$ acting on $\mathcal{M}$. To build the empirical operators, sparse N×N matrices $G_b$ were assembled such that

$$(G_b)_{ij} = P\left(x_i | x_j, b\right)$$

the conditional probability of reaching mode $x_i$ from $x_j$ via bitflip $b$, normalised with Laplace smoothing. The $G_b$ series forms the empirical generators of the discrete algebra. Algebraic structure was quantified in two ways. First, for each operator pair, the commutator

$$\left[M_i, M_j\right] = M_i, M_j \; - \; M_j, M_i$$

was evaluated, and the Frobenius norm $\|[M_i,M_j]\|$ was computed. Vanishing norms indicated commutativity.

Second, path algebra was computed via parsing successive triplet transitions to count ordered pairs ($M_i$, $M_j$). Asymmetries in counts between ($M_i$,$M_j$) and ($M_j$,$M_i$) indicated non-commutativity in practice—path-dependent operator orderings. Together, these analyses define a Lie-like discrete algebra over $\mathcal{M}$.

***Ricci Flow***

To analyse order and chaos in proteoform dynamics, each simulated trajectory was parsed into a time series of curvature values derived from occupancy $\rho(x)$ across the Boolean lattice $\mathcal{M}$ per the graph Laplacian. The resultant Ricci functionals quantified how curvature was distributed across the lattice. Two equivalent forms were used

$$\Phi(\rho) = \max_{x, y, \in \mathcal{M}} |R(x;\rho) - R(y;\rho)|$$

And

$$\Phi(\rho) = \sum_{x \in \mathcal{M}} \left(\rho_x - \frac{1}{|\mathcal{M}|}\right)^2$$

The first form measures the maximal curvature contrast across the lattice; the second is a variance-like form showing deviation of the occupancy distribution from flatness. Both return zero for a uniform distribution (globally flat manifold) and increase as curvature concentrates.

To map the spectrum of behaviours between order and chaos Ricci Flow was defined as bounded modal degeneracy

$$D(t) = \frac{1}{\Phi(t) + 1/N,} \; A(t)\frac{D(t) - 1}{N - 1}$$

where $N$ is the total number of modes. By construction, $D(t)$ is bounded between 1 (all molecules in one mode; maximal order) and $N$ (uniform distribution; maximal chaos). Hence, Ricci-flow dynamics can be expressed as a wave-like oscillation of degeneracy amplitude.

Bundles of $D(t)$ traces across replicate simulations were plotted to capture population-level dynamics. The ensemble mean and variance defined the central tendency and dispersion of degeneracy oscillations. To extract hidden oscillatory features, the Hilbert transform was applied to the ensemble-mean degeneracy trace, producing the analytic signal $z(t)$. From this, the oscillation envelope $|z(t)|$ (degeneracy amplitude) and the instantaneous frequency

$$v(t) = \frac{1}{2\pi}\frac{d}{dt}arg(x(t))$$

were derived. These quantities provide a spectral decomposition of Ricci-flow dynamics, with amplitude representing excursions between ordered and chaotic states, frequency capturing the cycling periodicity, and phase reflecting operator ordering along geodesic.

To characterise the oscillatory geometry of selected trajectories, they were projected into the complex domain via the analytical signal. For the Ricci spread series $\Phi(t)$, the projection was computed as

$$z(t) = Hilbert(\Phi(t) - \langle\Phi\rangle)$$

where the Hilbert transform returns the analytic signal with real and imaginary components corresponding to the in-phase and quadrature parts of the oscillation. This maps each Ricci trajectory to a continuous orbit in the complex plane, allowing visualization of symmetric versus trapped dynamics.

Selected trajectories were reduced to its dominant state sequence, from which two algebraic metrics were computed:

1. **Geodesic trap fraction**: the proportion of the trajectory that lies within a canonical geodesic path.
2. **Non-commutativity score**: the fraction of observed two-step bit-flip sequences that are inconsistent with any canonical geodesic ordering.

These metrics distinguish between *symmetric* trajectories (high trap fraction, low noncommutativity) and *trapped* trajectories (low trap fraction, high noncommutativity). Complex projection of these examples produced qualitatively distinct orbits: symmetric trajectories formed near-holomorphic loops, while trapped trajectories generated irregular, non-holomorphic paths with phase discontinuities.

***Conservation of Curvature***

In the single-molecule basis, the Boolean lattice $\mathcal{M}\{0,1\}^R$ was realised as the Hamming-1 graph with symmetric adjacency $W$ and Laplacian $L = D - W$. Curvature was defined as the discrete Laplacian $R = \Delta\rho = \Delta L$. Since $L^{\top}\mathbf{1} = 0$, total curvature vanishes for any occupancy

$$\sum_{x\in\mathcal{M}} R(x) = L^{\top} L\rho = 0,$$

Establishing a Noether-like discrete analogue: the volume-invariance of the $\mathcal{M}$ per axiom 2 enforced a curvature conservation law, as confirmed by computational instantiation confirming machine-zero sums for all single-site placements and random $\rho(x)$.

***The Geometric Primitive***

Atomic coordinates were obtained from the AlphaFold[1] model of human GAPDH (AF-P04406-F1, v4). Cysteine residues were identified, and their Cα coordinates extracted. Three cysteines (Cys152, Cys156, Cys247) were selected to define a 3-bit Boolean modal manifold ($R$=3).

To construct the substrate Laplacian, a weighted graph was constructed from all Cα atoms using a Gaussian kernel with cut-off $r_c =$ 6.5 Å and width $\sigma$ 4.0 Å. Edges between residues $i,j$ were weighted as:

$$w_{ij} = exp\left(-\frac{\|x_i - x_j\|^2}{2^{\sigma 2}}\right), \|x_i - x_j\| \leq r_c$$

From the weighted adjacency $W$, the combinatorial Laplacian was computed as $L = D - W$, where is $D$ the diagonal degree matrix.

To solve the PDE, each cysteine bit ($b$), a unit source term ($\rho_b$), was imposed at the Cα coordinate, and the discrete Poisson equation was solved

$$L\phi_b = \rho_b - \langle \rho_b \rangle$$

Solutions $\phi_b$were normalised to zero mean. The bitwise Dirichlet energy ($E_b$) and the local source energy ($S_b$) were computed as:

$$E_b = \frac{1}{2}\phi_b^{\mathrm{T}} L\phi_b, S_b = \frac{1}{2}\sum_j w_{bj}(\phi_b(b) - \phi_b(j))^2$$

The normalised pushforward weights were defined as

$$w_E b = \frac{S_b/E_b}{\sum_{\hat{b}} S_{\hat{b}}/E_{\hat{b}}}$$

To perform persistent topology analysis, local neighbourhoods of radius two edges were extracted from the substrate graph around each cysteine bit. Pairwise distances between neighbourhood residues were computed, and Vietoris–Rips filtrations applied to define two topological proxies:

- **H0 depth**: death time of the local cysteine bar (burial proxy).
- **H1 loopiness**: total persistence of finite H1 bars

Each proxy was normalised across bits, and combined as a morse-channel:

$$w_m b = \frac{depth_b / loop_b}{\sum_{\hat{b}}(depth_{\hat{b}} + Loop_{\hat{b}})}$$

The 3-bit modal manifold was represented as a Boolean hypercube, with the edges connecting Hamming-1 modes weighted according to their bitwise contributions. Multiple channel Laplacians were constructed: energy-only $(weights = w_E)$, topology-only $(weights = w_M)$, and a dual fused arithmetic mean $((w_E + w_M)/2)$ and tropical fusion, defined as:

$$w_{trop} b = \frac{\exp(-min(-log\ w_E(b), -log\ w_M(b)))}{\sum_{\hat{b}} \exp(-min(-log\ w_E(\grave{b}), -log\ w_M(\grave{b})))}$$

Dirichlet energy of the resultant modal Laplacian was evaluated by solving

$$L_d \phi = \rho - \langle \rho \rangle, \rho = \delta 000$$

and computing $E = \frac{1}{2} \phi^{\top} L_d \phi$. Then instantaneous transition probabilities from "000" were calculated as normalised edge weights. A probabilistic propagation of the modal manifold, was simulated under the master-equation kernel/ The distributions at selected times were compared across the Laplacians.

Curvature of the substrate graph was approximated using Forman-Ricci curvature. For each cysteine, curvature was computed as the mean edge curvature of its neighbours, rescaled to [0,1]. Curvature values were fused functorially with PDE-derived weights to define curvature-fused modal Laplacians.

***Numerical Example GAPDH Empirical***

The rate constant (*k*) for the reaction

$$RS^{-}(thiolate) + H_2O_2(hydrogen\ peroxide) \rightarrow RSOH\ (sulfenic\ acid)\ + H_2O$$

between Cys152 in GAPDH (2μM) and $H_2O_2$ (50μM) is $\approx 7M^{-1}s^{-1}$ *at* 20°C *and pH* 7.4, as calculated empricially[31], was calculated as the Gibbs free energy ($\Delta G^{\ddagger}$) activation energy per the Erying[31] equation

$$k = \kappa \frac{k_B T}{h} \frac{1}{C^{\circ}} exp(-\frac{\Delta G^{\ddagger}}{RT})$$

where $k_B$ is the Boltzmann constant, $h$ is Planck's constant, $T$ is temperature, $\kappa \approx 1$ is the transmission coefficient, and $C^{\circ} = 1M$ is the standard state concentration for bimolecular reactions. Rearranging,

$$\Delta G^{\ddagger} = -RT \ln(\frac{khC^{\circ}}{\kappa\ k_B T})$$

Substituting the empirical $k$ at 293.15 K yielded $\Delta G^{\ddagger} \approx 67\ kJ\ mol^{-1}$ $(16\ kcal\ mol^{-1})$.

**Curvature & Energy**

To estimate the activation energy using MGF Theory, the modal manifold was initialised. Allowed Hamming-1 edges were weighted by occupancy differences, and Forman-Ricci curvature was computed on each edge. This produced a symmetric curvature spectrum: each bit-flip from “000” carried the same curvature-derived barrier height. To give this curvature physical units, the magnitude of modal curvature was mapped to energy by:

$$\Delta G^{\ddagger}_{curv} = \kappa_{modal} \left| K_{ij} \right|,$$

Where $\kappa_{modal}$ is the geometric coupling constant ($J \cdot mol^{-1}$). This defines a baseline activation energy, that is symmetrical to each bit.

To differentiate this curvature bitwise, the real manifold projection weights $w_E(b)$ from the geometric primitive were applied as logarithmic scalars

$$\Delta G^{\ddagger}_{eff}(b) = \Delta G^{\ddagger}_{curv}(1 - In\ w_E(b))$$

This functorially lifts the PDE-derived heterogeneity into the modal manifold, producing differentiated barriers consistent with empirical kinetics. This approach preserves bitwise discriminatory power while inheriting physical units from the curvature–energy mapping.

***Scale-invariance***
To test scale-invariance, the source code used to implement the MGF Theory Field equation in the *R* = 3 basis was adapted to accommodate different *R* integers. The instantiated axioms and their evolution via the ALIVE algorithm remained invariant.

**Supplemental Information**

**Axiom 1 Extended to Multinomial Theorem.**

The structure of the entire proteoform state space ($\Omega_P$)—all theoretically possible molecular variants from a single copy of a UniProt accession or proteome comprising multiples thereof—can be derived from multinomial theorem:

$$|\Omega_P| = \sum_{PS=1}^{N_{PS}} (19^{L_{PS}}\; x\; M_{PS})$$

Where:

- $N_{PS}$ is the number of protein species (PS) inclusive of frameshifts, splice variants, and indels.
- $\sum_{PS=1}^{N_{PS}} is$ the sum for each PS.
- $19^{L,PS}$ is the genetic mutation state space for each AA, which can be mutated 19 ways from the original, for the PS, which can have different L due to indels.
- $M_{PS} = \prod_{i=1}^{L_{PS}} n_{i,PS}$ is the post-translational modification (PTM) state space for each AA in the PS, where $n_{i,PS}$ is the total number of PTM states (including unmodified) at residue $i$.

For example, a set comprising 10 PS with L = 100 where AA could be modified in at least one way, results in: $|\Omega_P|$ = $1.26\times10^{31}$ unique proteoforms. modes.

Multinomial theorem hierarchically organises this space Pascal Simplex ($\mathfrak{H}_P$). Within $\mathfrak{H}_P$, each proteoform occupies a unique coordinate in a multidimensional structure with two key planes:

1. **Horizontal Plane** – Encodes PTM state-space structure using multinomial coefficients, stratifying proteoforms by their modification-level degeneracy.
2. **Vertical Plane** – Encodes sequence-level divergence, placing near-identical PS (e.g., isoforms or mutations) closer together.

This multinomial Pascal simplex that structures the entire proteoform landscape across sequence variability and PTM states. All proteoform modes—whether arising from mutations, splicing, or PTMs—exist within a unified, mathematically constrained geometry. For example, GAPDH ($L$ = 335) exists within a multinomial coefficient space where its 111-cysteine mode is just one of 6,209,895 possible modes within the $k$-3 modification-stratified ensemble (0.89%-PTM-speciated) of 336-$k$-states.

This simplex demonstrates how combinatorics-enumerated modal state spaces can naturally be extended to accommodate different PTM types, such as sulfenic acids, disulfide bonds, and thiyl radicals in the cysteine basis.

**Axiom 2 Volume Invariance.**

On the modal Boolean manifold $X = \{0,1\}^R$ occupancy is described by a probability distribution $\rho_t(x)$. Conservation of total mass requires

$$\sum_{x \in \mathcal{M}} \rho_t(x) = \sum_{x \in \mathcal{M}} \rho_{t+1}(x) = 1$$

For a discrete update rule. Let $P_t(x \to y)$ be the probability of mode transition from *x* to *y*. Then

$$p_{t+1}(y) = \sum_{x \in \mathcal{M}} \rho_t(x) \, P_t(x \to y)$$

Conservation of probability holds if

$$\sum_{y \in X} P_t(x \to y) = 1 \; \forall x \in X$$

That is, $P_t$ is a stochastic matrix. To define the continuity equation in flux form. Let $J_t(x \to y)$ $=\rho_t(x) \, P_t(x \to y)$ then

$$\rho_{t+1}(x) - \rho_t(x) = \sum_{y \in N(x)} [J_t(y \to x) - J_t(x \to y)]$$

where *N(x)* are the Hamming-1 neighbours of *x*. Summing over all *x* cancels fluxes and guarantees conservation.

In continuous time series with transition rates $k_{xy} \geq 0$

$$\dot{\rho}(x) = \sum_{y e N(x)} \left[\rho(y) k_{yx} - \rho(x) k_{xy}\right]$$

with $\sum_x \dot{\rho}(x) = 0$.

The dynamic synthesis and degradation of instantiated molecules is accommodated by the source term $s_t(x)$

$$\rho_{t+1}(x) - \rho_t(x) = \sum_{y \in N(x)} [J_t(y \to x) - J_t(x \to y)] + s_t(x)$$

With total occupancy

$$N_{t+1} = \sum_x \rho_{t+1}(x) = N_t + \sum_x s_t(x)$$

One obtains

$$\rho_{t+1}(y) = \sum_x \rho_t(x) P_t(x \to y)$$

Since both numerator and denominator scale with $N_t$, the stochastic $P_t$ kernel remains invariant regardless of the total occupancy of the $\mathcal{M}$, and the normalised probability distribution continues to satisfy $\sum_x \rho_t(x) = 1$.

Hence, the volume of the modal manifold (sum of possible modes) is invariant: no change is permitted, so the result of any transformation (redistribution of molecules) is exactly zero. When a set of molecules occupy the manifold, their occupancies are always parts of a whole that sum to one—whether expressed as a probability distribution or as an absolute total—and this remains invariant even under synthesis or degradation, once renormalised.

**Non-symplectic Modal Manifold**
***Setting***: The modal manifold $\mathcal{M}$ is $\{0,1\}^R$ finite and discrete. Modes are the vertices; allowed transitions use Hamming-1 edges. Any metric $g$ and curvature $R[g]$ are defined on this graph (e.g., via Laplacians/Forman–Ollivier), not by a smooth structure.

***Proposition of no symplectic form on $\mathcal{M}$***. A symplectic structure on a manifold $\mathcal{M}$ is a closed, non-degenerate 2-form $\omega \in \Omega^2(M)$. For each $x \in M$, nondegeneracy requires the bilinear map:

$$v \mapsto \iota_v \omega_x : T_x M \to T_x^* M$$

to be an isomorphism.

For the discrete modal manifold, each tangent space is trivial

$$T_x M = \{0\} \; \forall x \in \mathcal{M}$$

Hence $\Lambda^2 T_x^* \mathcal{M} = 0$ and every 2-form vanishes identically: $\omega \equiv 0$. The zero 2-form, therefore, is degenerate. Hence, the $\mathcal{M}$ is non-symplectic.

**Remark 1**: Endowing $\mathcal{M}$ with any edge-weighted metric $g$ and computing discrete curvature $R[g]$ (via graph Laplacians, Forman/Ollivier, etc.) does not change the underlying differentiable structure: $T_x M$ remains $\{0\}$. Hence, $\Omega^2(M) = \{0\}$, meaning no nondegenerate closed 2-form can arise. The manifold remains non-symplectic.

**Remark 2**: Volume invariance in Axiom 2 is therefore combinatorial (probability-mass conservation under a stochastic update / conservative generator), not a consequence of Liouville's theorem on a symplectic phase space.

**Remark 3**: Sometimes the modal manifold is represented as a probability simplex

$$\Delta^{N-1} = \left\{ \rho \in \mathbb{R}^N : P_i \geq 0, \sum_i P_i = 1 \right\}$$

However, $\Delta^{N-1}$ does not admit a **canonical symplectic structure**:

- A symplectic form requires a nondegenerate, closed 2-form ω, which is only defined on smooth even-dimensional manifolds.

- While $\Delta^{N-1}$ is smooth, the dynamics of $\rho$ are Markovian (given by stochastic matrices or conservative generators). Markov flows are dissipative (entropy can increase), not Hamiltonian, and therefore cannot be written as symplectic flows.
- Any attempt to endow $\Delta^{N-1}$ with a symplectic form is artificial and not preserved by the dynamics.

Hence, whether one views the modal manifold as the discrete Boolean lattice or its embedding into the probability simplex, the system remains fundamentally non-symplectic. Conservation of occupancy arises from combinatorial structure and stochastic conservation laws, not Hamiltonian mechanics.

**Axiom 3 First Order Dynamics.**
If time or space is fundamental, then each reaction corresponds to a distinct event that must be localised either temporally or spatially. Such events cannot be perfectly simultaneous. Special relativity shows that absolute simultaneity does not exist; the ordering of spatially separated events depends on the frame of reference. At the molecular scale, this means that what appears as a “two-site jump” is not a primitive transition but rather the composition of two sequential first-order events, however, finely separated in time or space.

This principle is entirely consistent with calculus: continuous processes are understood as limits of infinitesimal increments. On a discrete manifold, the analogue is that multi-site modal changes ($\Delta k$>1) necessarily decompose into sequential first-order transitions along the Boolean edges of the modal hypercube.

***Geometric corollary:*** Formally, $\mathcal{M} = \{0,1\}^R$ is the $R$-dimensional Boolean hypercube, with vertices corresponding to modes and edges corresponding to single-site transitions. This can be represented as a binomial diamond, by stratifying modes according to Hamming weight ($k$). This projection preserves the combinatorial structure relevant to modal dynamics.

Hence, on the Boolean manifold, all transitions with Δk>1 necessarily decompose into sequential Hamming-1 steps. First-order dynamics are therefore a fundamental invariant of modal geometry.

**Axiom 4 Occupancy Curves the Modal Manifold.**
***Categories and functorial projection***. Real and abstract are coupled via the functor:

Real category $\mathcal{C}_{REAL}$

- **Objects**: pairs $(\rho, \varphi)$, where $\rho \in \Delta^{N-1}$ is the occupancy distribution on the modal manifold $\mathcal{M} = \{0,1\}^R$, and φ encodes molecular shape (e.g., atomic structure).
- **Morphisms**: Conservative updates $\rho \mapsto P\rho$ with column-stochastic $P$, optionally including source/sink terms for open systems (synthesis, degradation).

Abstract category $\mathcal{C}_{ABS}$

- **Objects**: pairs $(g, R)$, where g is a metric on $\mathcal{M}$ and $R: \mathcal{M} \rightarrow \mathbb{R}$ is the associated curvature field.
- **Morphisms**: induced updates $(g, R) \;\mapsto\; (\grave{g}, \grave{R})$ under occupancy redistribution.

Functor

$$\mathcal{F}: \mathcal{C}_{REAL} \longrightarrow \mathcal{C}_{ABS}, \qquad \mathcal{F}(\rho, \varphi) = (g(\rho, \varphi), R(\rho, \varphi))$$

Functoriality ensures that composition is preserved

$$\mathcal{F}(f \circ g) = \mathcal{F}(f) \circ \mathcal{F}(g)$$

so any sequence of real molecular updates induces a lawful algebra of transformations in the abstract geometric category.

***Metric and curvature conservation.***
**Scalar field**. Construct a scalar field

$$\sigma = S(\rho, \varphi)$$

where $\rho$ encodes occupancy and $\varphi$ encodes shape. This can be a linear combination, smoothing, or PDE solution on the contact-map Laplacian.

**Metric**. Define a conformational deformation of the base metric $g_0$:

$$g(\rho, \varphi) = e^{2\sigma} g_0$$

**Curvature operator**. Let $L_\varphi$ be a symmetric zero-sum Laplacian defined by shape $\varphi$. Define curvature as

$$R(x) = (L_\varphi \sigma)(x).$$

**Conservation of curvature**. Because $\mathbf{1}^{\intercal} L_\varphi = 0$,

$$\sum_{x \in \mathcal{M}} R(x) = \mathbf{1}^{\intercal} L_\varphi \sigma = 0$$

Hence, curvature is conserved globally: it cannot be created or destroyed, only redistributed across $\mathcal{M}$.

Under any update $\grave{\rho} = P\rho$,

$$\grave{R} = L_\varphi \grave{\sigma}, \sum_{x} \grave{R}(x) = 0$$

Hence, conservation holds for all modal transitions.

**Geometric corollaries**

- **Differentiation of dynamics:** Curvature deforms the manifold, meaning that transitions with equal energy cost can nonetheless be distinguished by geometric penalty.
- **Energy coupling:** In the transition law, curvature introduces an additive term that acts as a geometric contribution to activation energy.

- **Breaking iso-potential symmetry:** Without curvature, certain transitions would be iso-potential and indistinguishable; with curvature, they are differentiated.

Hence, curvature is the **conserved shadow of molecular identity**: projected from reality, encoded geometrically, and transported under all transformations.